# Enhancing second harmonic generation using dipolar-parity modes in non-planar plasmonic nanocavities

Feng Wang,[1] Manoj Manjare,[1] Robert Lemasters,[1] Chentao Li,[1] Hayk Harutyunyan[1,*]

[1]Department of Physics, Emory University, Atlanta, 30322 GA, USA
*Corresponding author: hayk.harutyunyan@emory.edu

**There is an active demand to develop efficient nanoscale nonlinear sources for applications in photonic circuitry, quantum optics and biosensing. To this end, plasmonic systems have been utilized to boost the nonlinear signal generation, however high efficiencies of frequency conversion for realistic applications remain a challenge. Metal-insulator-metal (MIM) nanocavities are good candidates for strongly concentrating the fields at the nanoscale to enhance the optical nonlinearities, however they typically suffer from the requirement to have a quadrupolar resonance at the emission wavelength. Here, we introduce nonplanar MIM nanocavities with a nonlinear spacer that can strongly enhance the second harmonic generation (SHG) despite of having fundamental and emission modes of the same parity. Our experimental and numerical results indicate that the enhancement is due to the non-planar design of the cavities and the bulk nonlinearity of the spacer layer.**

The emergence and the rapid advance of nanophotonic systems allowed an unprecedented control of the optical modes and interactions at the nanoscale [1]. This has been especially beneficial for nonlinear optical processes as the subwavelength nature of these platforms has opened up a range of novel possibilities [2, 3]. For example, at the nanoscale the strict phase matching conditions are relaxed, unlocking a host of strongly nonlinear materials that cannot be used in the bulk [4-6]. Furthermore, the field confinement afforded by nanoresonators concentrates and enhances the local fields leading to greater increase in the efficiency of the nonlinear light-matter interactions [2]. Finally, the ability to control the spatial distribution and symmetry of optical modes at the nanoscale is important for the engineering of strong nonlinearities [7, 8].

In this context, plasmonic systems have been used extensively to confine the optical fields into deep subwavelength volumes and to enhance the nonlinear optical phenomena at the nanoscale. Various nonlinear processes have been shown to benefit from plasmonic enhancement including SHG, third harmonic generation, four wave mixing, optical Kerr effect [9-22]. MIM nanocavities are of particular interest as they can reproducibly achieve strong field enhancement in the thin insulating layer [23, 24]. Moreover, the plethora of modes supported by MIM nanoresonators allows to design multi-resonant structures enhancing the optical fields at all the frequencies involved in the nonlinear process [10].

2nd order processes such as SHG require inversion symmetry breaking for electron motion. The pure plasmonic structures are made of noble metals which possess centrosymmetric crystal lattice and thus the symmetry breaking is achieved at the material interface. Thus, the effective nonlinear susceptibility tensor element $\chi^{(2)}_{\perp\perp\perp}$ is typically the dominant source for nonlinearity [25]. The major drawback of this approach is that it requires the optical mode at the SHG frequency to have different parity than the fundamental mode [7]. This requirement stems from the fact that the nonlinear emission detected in the far field is a surface integral of the local nonlinear polarizability and the SHG mode. Thus, dipolar-like distribution for both fundamental and SHG modes would interfere destructively in the far field resulting in zero SHG [13]. This leads to the necessity of having even-parity modes such as quadrupoles at the SHG wavelength which may not be highly radiative due to the poor coupling to the free space radiation or not accessible because of the gap plasmon dispersion [10].

An alternative approach for using plasmonics for efficient frequency conversion is to couple non-plasmonic, highly nonlinear materials to metallic nanoparticles [26-28]. In such a scheme, plasmonic near fields concentrate light inside the adjacent nanomaterial thus generating strong nonlinearities in those materials. In these hybrid systems the SHG is mainly due to the bulk nonlinearity of the non-plasmonic material and is not dominated by the symmetry breaking at the metal interface.

Here, we use hybrid plasmonic-nonlinear dielectric nanocavities to design doubly resonant platform that does not rely on quadrupolar modes for observing SHG. We show that if the signal predominantly originates from the nonlinear dielectric, the nanocavities can be designed such that both fundamental and SHG resonance can have dipole-like parities.

The proposed MIM nanocavities are shown in Fig. 1a. They comprise of a Au nanowire grating separated from a thick silver film by a thin ZnO spacer layer that acts as a nonlinear material. The Wurtzite structure of ZnO possesses $C_{6v}$ symmetry with nonvanishing $\chi^{(2)}_{zzz}$ element, which is ideal for the fields polarized normal to the metal surface in MIM cavities [29]. The Au nanowires are fabricated on a cleaned SiO$_2$ substrate by -beam lithography followed by thermal evaporation of gold. The designed widths of Au nanowires vary between 60 and 160 nm, whereas their thickness is fixed at 35 nm. To study the effect of grating periodicity, samples with four different periods (420 nm, 440 nm, 460 nm and 480 nm) are fabricated. A representative SEM of the fabricated grating (140 nm width, 420 nm period) is shown in Fig. 1b. The grating fabrication is followed by atomic layer deposition (ALD) of 12.5 nm ZnO spacer layer which conformally covers the entire surface of Au nanowires including their sides. As a final step, 150 nm thick Ag film is deposited on top of the ZnO layer, to serve a second metal layer. In the resulting MIM design, non-planar plasmonic nanocavities are formed between the Au nanowires and Ag film. The details of optical characterization setups have been published elsewhere [10]. Briefly, the SHG experiments are carried out using 150 fs, 80 Mhz pulses from a wavelength-tunable Optical Parametric Oscillator and high numerical aperture (NA=1.4) lens. A small beam diameter (~2 mm), compared to the large back aperture diameter of the objective (11mm), is used to ensure near-normal excitation.

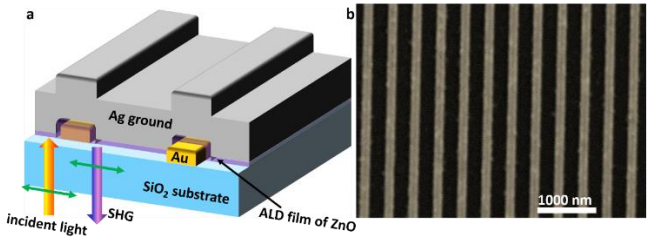

Fig. 1. (a) Schematic of the proposed non-planar patch nano-antennas grating; (b) representative SEM of the fabricated gratings.

First, we characterize the nanocavities in the linear regime to identify their resonances. For this, we measure reflection spectra using normally-incident white-light excitation through low NA=0.25 lens. The polarization of the incident light is perpendicular to the grating nanowires. The measured visible and near-infrared (NIR) spectra for different nanowire widths with a fixed 440 nm grating period are shown in Fig. 2. The visible range (Fig. 2a) features two distinct dips (Modes 2 and 3), whereas a broad dip (Mode 1) is observed in NIR (Fig. 2b). All of these resonances experience various degrees of red-shift with increasing nanowire widths. In contrast to this, only the Mode 2 shifts to the longer wavelengths with increasing the periodicity of the gratings from between 420 nm to 480 nm (Fig. 2c and 2d). Thus, the period-dependent mode is identified as a hybridized mode between the propagating plasmons on the Ag surface and gap modes of non-planar cavities [10], whereas the period-independent modes are pure cavity modes of the MIM system. The latter two modes are the dipolar and hexapolar modes that will be used for doubly-resonant SHG.

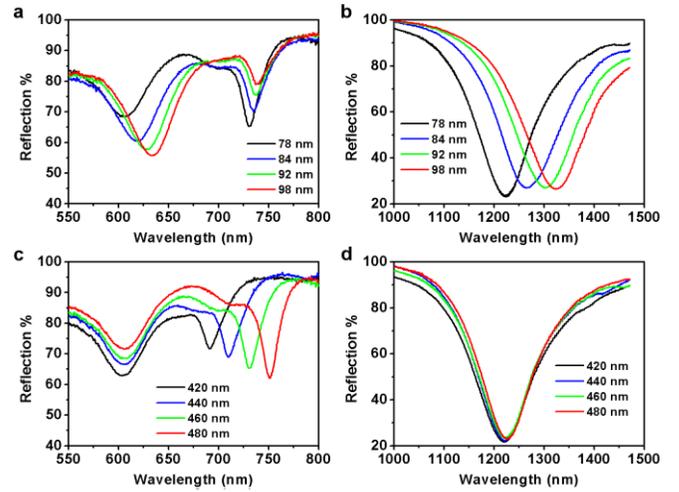

Fig. 2. Experimental reflection spectra. (a) visible, and (b) NIR spectra for fixed period 440 nm and varying nanowire widths between 78 nm and 98 nm. c) visible, and d) NIR spectra for fixed nanowire width 78 nm and varying period between 420 nm and 480 nm.

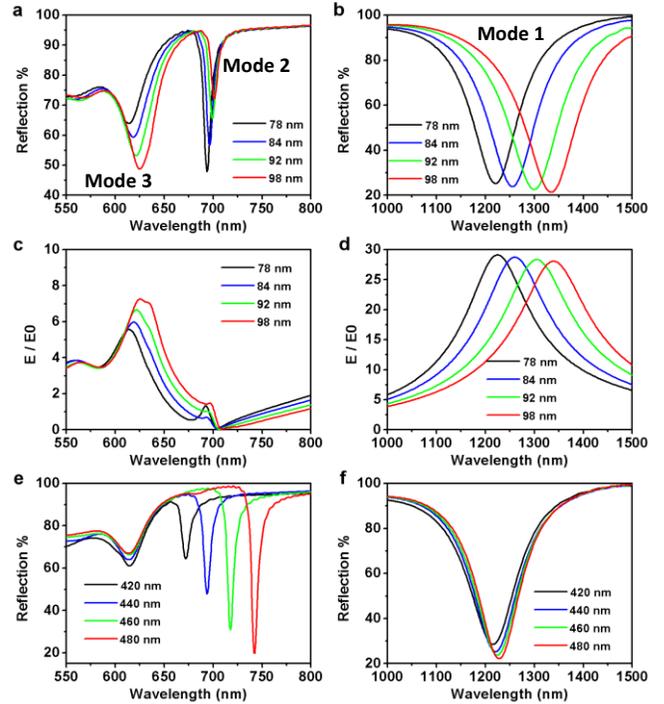

Fig. 3. Simulated spectra. (a) visible and (b) NIR reflection spectra for a fixed nanowire period at 440 nm and varying widths between 78 nm and 98 nm. (c) visible and (d) NIR local field enhancement spectra for fixed period at 440 nm and varying widths between 78 nm and 98 nm. (e) visible and (f) NIR reflection spectra for fixed nanowire width at 78 nm and varying period between 420 nm and 480 nm.

In order to understand the measured reflection curves in more detail, we perform numerical simulations. The geometric parameters used in the simulations are taken from the fabricated samples shown in Fig. 2. Based on the simulated field distribution in the cavities we calculate the wavelength-dependent Poynting vectors and then extract the reflection spectra.

Fig. 3a and 3b show the calculated reflection curves for a fixed period (440 nm) and varying nanowire widths (78 nm to 98 nm). The shape and the trend of these curves are in a good agreement with the measured spectra. Next, we calculate the near field enhancement spectra for the same samples and confirm that the reflection dips correspond to local field enhancement, i.e. the resonant modes of the system (Fig. 3c and 3d). The spectra for fixed width and varying period shown in Fig. 3e and 3f further confirm that the only periodicity-dependent resonance is Mode 2, whereas Modes 1 and 3 are pure cavity modes.

After determining the resonant wavelengths of the cavity modes, we perform SHG experiments on the MIM samples by sweeping the laser pump wavelength between 1100 nm and 1500 nm under near-normal excitation conditions. In order to determine the SHG enhancement and to avoid any wavelength-dependent detection efficiency artifacts, the measured nonlinear signal at each excitation wavelength was normalized by the SHG signal obtained from a calibration sample (12 nm ZnO film deposited on a 150 nm smooth Ag film). Fig. 4a shows the spectral dependence of the SHG enhancement factors for different nanowire widths with a fixed period (440 nm). Considering the previously-determined wavelengths of the cavity resonances, there are several features to note in Fig. 4a. First, the pump wavelengths corresponding to the maximum SHG enhancement do not coincide with the center wavelengths of the fundamental modes (Mode 1), although they do shift to longer wavelengths with increasing widths of the nanowires. In fact, the peaks of SHG enhancement are more consistent with the wavelengths of the higher order cavity modes, i.e. the Mode 3 in Fig. 2 and 3. Thus, consistent with the existing literature, our data indicates that the emission wavelength resonance plays an important role for the enhancement of the nonlinear signal. Second, the maximum SHG enhancement values gradually decrease with increasing the width of the nanowires. The SHG is the strongest for 78 nm width, with its peak values measured at 1200 nm. The reason for this is that the resonant wavelength of the fundamental modes are very close to the double of the wavelength of the higher order cavity resonance. Due to the strongly sub-linear dispersion of gap plasmonic mode [10], this wavelength matching condition deteriorates for larger widths leading to diminished a SHG. To make this point clearer, in Table 1 we list the wavelength mismatch between Mode 1 and Mode 3 and the local field enhancement due to this mismatch.

**Table 1. Comparison of resonant wavelengths and mode mismatch in 440 nm period gratings.**

| Nanowire width | 78 nm | 84 nm | 92 nm | 98 nm |
|---|---|---|---|---|
| Mode 1 wavelength $\lambda_1$ | 1220 nm | 1260 nm | 1301 nm | 1323 nm |
| Mode 3 wavelength $\lambda_3$ | 605 nm | 619 nm | 627 nm | 631.5 nm |
| Wavelength mismatch $\frac{\lambda_1}{2} - \lambda_3$ | 5 nm | 11 nm | 23.5 nm | 30 nm |
| Cavity field enhancement at $\frac{\lambda_1}{2}$ (simulated) | 5.7 | 4.75 | 3.4 | 3.15 |

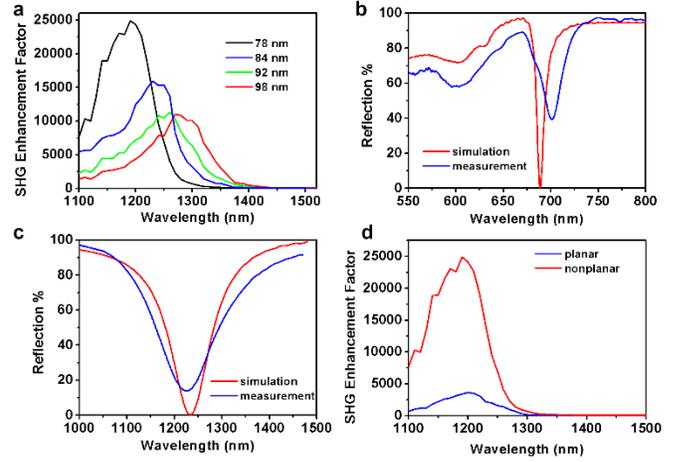

Fig. 4. (a) measured SHG enhancement factors of non-planar nanocavities with different nanowire widths and fixed 440 nm grating period; (b) visible and (c) NIR reflection spectra of planar nanocavities (440 nm period, 110 nm nanowire width and 8 nm ZnO thickness); (d) comparison of SHG enhancement factors between planar and nonplanar nanocavities, with both samples having fundamental modes at ~1220 nm and higher order modes at ~610 nm.

To further understand the origin of SHG enhancement we fabricate control samples which feature planar MIM nanocavities. Here, Au gratings with 110 nm width and 20 nm thickness are defined on top of a 150 nm thick Ag film covered with 8 nm thick ZnO spacer layer. The periodicity is fixed at 440 nm. The sample is covered with a thick index-matching oil layer for optical measurements. For a reflection measurement we observe spectra similar to the nonplanar samples, with Mode 1 at ~1200 nm, Mode 2 at ~700 nm and Mode 3 at ~600 nm (Fig. 4b and 4c). For SHG enhancement spectra (Fig. 4d), we observe an order of magnitude smaller signal compared to the nonplanar design.

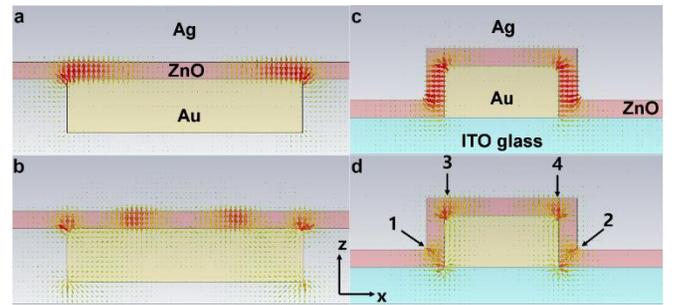

Fig. 5. (a) and (b) mode field distribution at 1220 nm and 610 nm for planar nanocavities (period 440 nm, nanowires width 110 nm, ZnO film thickness 8 nm); c) and d) mode field distribution at 1220 nm and 610 nm for non-planar patch antennas (period 440 nm, nanowires width 78 nm, ZnO film thickness 12 nm).

The SHG enhancement data suggests that the increase in nonlinear generation is supported by the doubly-resonant nature of our samples. However, in striking difference to the existing literature both of these resonances are dipolar in nature. Moreover, the comparison between the planar and nonplanar samples suggests that the latter design is critical for the observation of strong SHG. We argue that this is the result of the fact that the nonlinear signal originates in ZnO film rather than at the metal interface.

According to nonlinear scattering theory, the SHG in the far field $E_{SHG}$ can be written as [30]

$$E_{SHG} \propto \int \boldsymbol{P}_{NL} \cdot \boldsymbol{E}(2\omega)\, dV, \qquad (1)$$

where $\boldsymbol{P}_{NL}$ is the nonlinear polarization excited by the pump and $E(2\omega)$ is the local mode at the SHG frequency excited from far field. For plasmonic systems where the SHG occurs due to the symmetry breaking at the metal interface Eq. 1 reduces to a surface integral

$$E_{SHG} \propto \int \chi^{(2)}_{\perp\perp\perp} E_\perp(\omega) E_\perp(\omega) E_\perp(2\omega)\, dS, \qquad (2)$$

where $E_\perp(\omega)$ and $E_\perp(2\omega)$ are the electric field components normal to the metal surface for local modes at the fundamental and SHG frequencies, respectively. According to Eq. 2 if the fundamental and emission modes are of the same dipolar parity the nonlinear signal at the detector will destructively interfere resulting in no SHG. However, if the source of the SHG is the bulk nonlinearity of the ZnO then the value of the volume integral in Eq. 1 will depend on the geometrical shape of the nanocavity and can be written as

$$E_{SHG} \propto \int \overleftrightarrow{\chi}^{(2)} E(\omega) E(\omega) E(2\omega)\, dV, \qquad (3)$$

where $\overleftrightarrow{\chi}^{(2)}$ is the ZnO bulk nonlinear susceptibility tensor. To understand why the integral in Eq. 3 will not vanish we study the local field distributions of the fundamental and emission modes for planar and nonplanar designs, Fig. 5. The fundamental modes are dipolar (so-called magnetic dipole modes in MIM structures) whereas the SHG modes possess 3 wave nodes, i.e. have the same parity as the fundamental mode. For planar system the local fields at the opposite ends of the nanocavity point in opposite directions due to the dipolar nature of the resonances for both excitation and emission wavelengths (Fig. 5a and 5b). This leads to the emission cancellation in the far field as the integral in Eq. 3 is effectively zero similar to Eq. 2. On the other hand, for the nonplanar geometry the fields of the fundamental mode (Fig. 5c) at the opposite sides of the nanocavity are polarized in the same direction along the $x$ axis and so are the fields at regions 1 and 2 for the emission mode (Fig. 5d). Thus, the integral in Eq. 3 does not vanish leading to an enhanced SHG in the far field. The fields in regions 3 and 4 in Fig. 5d would still cancel out due to the destructive interference. In other words the difference between Eq. 2 and 3 lies in the nature of the nonlinearity. In the case of dominant surface nonlinearity the fields are defined with respect to the normal to the metal surface, rendering the surface integral zero, whereas in the case of bulk nonlinearity the fields are defined with respect to global coordinate system and thus interfere constructively in the case of the nonplanar design.

In conclusion, we show that multi-resonant MIM cavities can be utilized to enhance the SHG at the nanoscale. Due to the gap plasmon dispersion, it is typically challenging to design MIM cavities with quadrupolar mode wavelengths that match the double wavelength of the fundamental modes. In this work we use 3$^{rd}$ order modes to match the SHG wavelength. By using the bulk nonlinearity of the spacer layer we show that such systems can exhibit strong far field SHG despite having the same parity for both excitation and emission modes. The resulting efficient SHG nanosource will pave the way for novel integrated photonic applications and nonlinear devices.

**Funding.** National Science Foundation (NSF)(EFMA – 1741691); Research Corporation for Science Advancement (24371).